\documentclass[twocolumn,showpacs,prl]{revtex4}
\usepackage{amssymb}
\usepackage{amsmath}
\usepackage{graphicx}
\usepackage{dcolumn}
\usepackage{bm}

\begin{document}

\title{Experimental realization of an entanglement filter through the
environmental selection}
\author{C. Zu$^{1}$, Y.-X. Wang$^{1}$, X.-Y. Chang$^{1}$, P.-Y. Hou$^{1}$,
H.-X. Yang$^{1}$, L.-M. Duan$^{1,2}$}
\affiliation{$^{1}$Center for Quantum Information, IIIS, Tsinghua University, Beijing,
China}
\affiliation{$^{2}$Department of Physics, University of Michigan, Ann Arbor, Michigan
48109, USA}

\begin{abstract}
We report an experiment that uses the environmental selection, a key concept
in the recent theory of quantum Darwinism, as a mechanism to realize the
entanglement filter, a useful quantum information device that filters out
certain entangled states. In the experiment, the environment of two qubits
is controlled to favor an entangled state and kill other competing
components in the input state. The initial state has vanishing entanglement,
but the state surviving after interaction with the environment is close to a
maximally entangled state, with an entanglement fidelity of $\left( 94.7\pm
1.9\right) \%$\ measured through the quantum state tomography. We
experimentally demonstrate that the generated entanglement is robust under change of
the initial state configurations and the environmental parameters.
\end{abstract}
\maketitle

Environmental selection, a key concept in Darwin's theory of evolution, has
found important applications in several areas of science. A striking example
in quantum physics is the mechanism known as quantum Darwinism, recently
proposed to explain the emergence of classical reality from the quantum
world \cite{1,2,3,4,5}. The concept of quantum Darwinism has found
interesting applications in theory of quantum measurements \textbf{\cite%
{1,2,3,4,5}}. In quantum measurement process, the fittest states selected
out by the environment correspond to the classical pointer states \textbf{%
\cite{1,2}}. The state of the system, originally in a quantum superposition,
tends to collapse onto the pointer states due to the environmental
interaction. The proliferation of the classical pointer states in the
environment explains their robustness and the appearance of a classical
reality through the measurement, but at the same time it induces decoherence
in the system, reducing its potential for quantum information. \textbf{\cite%
{1,2}}.

With advance of quantum control techniques \cite{6}, it becomes possible to
engineer the environment of quantum systems in such a way that the fittest
state selected out by the environment corresponds to a quantum entangled
state. The environmental selection, a key concept in quantum Darwinism, can
then be used in a novel way as a mechanism to realize an entanglement
filter. Just like a polarization filter which can select out certain
polarization states, the entanglement filter is a useful quantum information
device proposed in Refs. \cite{6a,6b} to filter out certain entanglement
components. An entanglement filter that can select out two-qubit states
according to the parity of their polarizations has been realized recently
\cite{6b}.

In this paper, we report an experiment that demonstrates a new type of
entanglement filter based on the mechanism of the environmental selection.
This entanglement filter selects out a certain entangled Bell state, which
corresponds to the fittest state in a specially designed environment for two
propagating photonic qubits. Through the environmental selection, the
entanglement filter generates almost maximally entangled output state with a
measured entanglement fidelity about $\left( 94.7\pm 1.9\right) \%$ from the
unentangled input states. Note that the environmental coupling is typically
in favor of classical states and leads to decay of quantum entanglement.
Dissipation from the environmental coupling has been generally identified as
the major obstacle for realization of quantum information processing. There
are striking exceptions where controlled dissipation under certain
configurations become helpful for entanglement generation \textbf{\cite%
{7,8,9,10,11,12}}. Our reported experiment belongs to this striking class,
where the environmental selection emerging from the dissipation is used as a
mechanism to realize an entanglement filter, which generates entanglement
from unentangled input states.

In our experiment, the input state is a mixture of two classical states%
\begin{equation}
\rho _{in}=\left\vert c_{0}\right\vert ^{2}\left\vert HV\right\rangle
\left\langle HV\right\vert +\left\vert c_{1}\right\vert ^{2}\left\vert
VH\right\rangle \left\langle VH\right\vert
\end{equation}%
with arbitrary mixing coefficients $\left\vert c_{0}\right\vert ^{2}$ and $%
\left\vert c_{1}\right\vert ^{2}$, where $\left\vert H\right\rangle $ ($%
\left\vert V\right\rangle $) represents a single-photon state with
horizontal (or vertical) polarization. The input state $\rho _{in}$ is in a
two-dimensional subspace spanned by $\left\vert HV\right\rangle $ and $%
\left\vert VH\right\rangle $. Alternatively, we can take the Bell states $%
\left\vert \Psi ^{\pm }\right\rangle =\left( \left\vert HV\right\rangle \pm
\left\vert VH\right\rangle \right) /\sqrt{2}$ as the basis-vectors for this
subspace. Using $\left\vert \Psi ^{\pm }\right\rangle $ as the
basis-vectors, the state $\rho _{in}$ is expressed as $\rho _{in}=\left(
\left\vert \Psi ^{+}\right\rangle \left\langle \Psi ^{+}\right\vert
+\left\vert \Psi ^{-}\right\rangle \left\langle \Psi ^{-}\right\vert \right)
/2+\left( \left\vert c_{0}\right\vert ^{2}-\left\vert c_{1}\right\vert
^{2}\right) \left( \left\vert \Psi ^{+}\right\rangle \left\langle \Psi
^{-}\right\vert +\left\vert \Psi ^{-}\right\rangle \left\langle \Psi
^{+}\right\vert \right) $. We design an environment so that it favors a
certain entangled state, say $\left\vert \Psi ^{+}\right\rangle $, and leads
to large decay of the other component ($\left\vert \Psi ^{-}\right\rangle $
in this case) with a decay ratio $\beta =e^{-\gamma _{-}}/e^{-\gamma _{+}}<1$%
. This realizes an entanglement filter that selects out the $\left\vert \Psi
^{+}\right\rangle $ state. After the environment, the effective output state
(unnormalized) has the form%
\begin{eqnarray}
\rho _{out} &=&\left( \left\vert \Psi ^{+}\right\rangle \left\langle \Psi
^{+}\right\vert +\beta \left\vert \Psi ^{-}\right\rangle \left\langle \Psi
^{-}\right\vert \right) /2  \notag \\
&+&\left( \left\vert c_{0}\right\vert ^{2}-\left\vert c_{1}\right\vert
^{2}\right) \sqrt{\beta }\left( \left\vert \Psi ^{+}\right\rangle
\left\langle \Psi ^{-}\right\vert +\left\vert \Psi ^{-}\right\rangle
\left\langle \Psi ^{+}\right\vert \right) .
\end{eqnarray}%
In the limit with $\beta \ll 1$, the output state becomes maximally
entangled although the input state $\rho _{in}$ apparently has no
entanglement. In the simple case of $\left\vert c_{0}\right\vert
^{2}=\left\vert c_{1}\right\vert ^{2}$, the entanglement of $\rho _{out}$,
measured by the concurrence $C$ \textbf{\cite{13}}, has the analytic form $%
C=\left( 1-\beta \right) /\left( 1+\beta \right) $.

To experimentally produce the mixed state in the form of Eq. (1) for two
photons, we generate the photon pair through the spontaneous parameter down
conversion (SPDC) in a nonlinear BBO\ crystal. The experimental setup is
shown in Fig. 1. The pumping pulse before the BBO crystal is in the
polarization state $c_{0}\left\vert H\right\rangle +c_{1}\left\vert
V\right\rangle $, where the ratio $c_{0}/c_{1}$ is set by the angle of a
half wave plate (HWP). After the BBO\ crystal, the down converted photons
are in the state $\left\vert \Psi \right\rangle =c_{0}\left\vert
H,f_{H}\left( t\right) \right\rangle \left\vert H,f_{H}\left( t\right)
\right\rangle +c_{1}\left\vert V,f_{V}\left( t\right) \right\rangle
\left\vert V,f_{V}\left( t\right) \right\rangle $, where $\left\vert
f_{H}\left( t\right) \right\rangle $ ($\left\vert f_{V}\left( t\right)
\right\rangle $) denotes the pulse shape of the corresponding photon with
horizontal (vertical) polarization. Due to the birefringence in the BBO\
crystal, the shape functions $f_{H}\left( t\right) $ and $f_{V}\left(
t\right) $ for different polarization components do not perfectly overlap in
time. To generate entanglement, typically we need another birefringent
crystal, such as a quartz, to compensate this mismatch in the pulse shapes.
In this experiment, we set the quartz compensator in the reserve direction,
so the shape mismatch of the pulses is actually amplified. As a result, we
have $\left\langle f_{H}\left( t\right) \right. \left\vert f_{V}\left(
t\right) \right\rangle $ $\approx 0$, and the polarization state of the down
converted photons, after tracing over the shape degrees of freedom, is
described by the mixed state $\left\vert c_{0}\right\vert ^{2}\left\vert
HH\right\rangle \left\langle HH\right\vert +\left\vert c_{1}\right\vert
^{2}\left\vert VV\right\rangle \left\langle VV\right\vert $. After another
HWP which exchanges the states $\left\vert H\right\rangle $ and $\left\vert
V\right\rangle $ for one of the photons, we get a state in the form of $\rho
_{in}$ described by Eq. (1).

\begin{figure}[tbp]
\includegraphics[width=8cm,height=9cm]{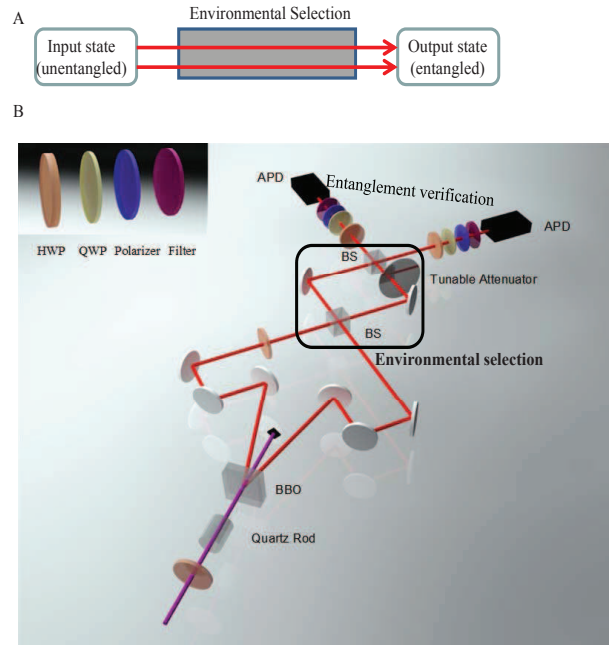}
\caption[Fig. 1 ]{ (A) Illustration of an entanglement filter through the
environmental selection. (B) Experimental setup that realizes an entanglement filter
through the environmental selection. Before the box, a
spontaneous parametric down conversion process prepares two single photons
in unentangled states. Ultrafast laser pulses (with the pulse duration less
than $150$ fs and a repetition rate of $76$ MHz) at the wavelength of $400$
nm from a frequency doubled Ti:sapphire laser pump two joint
beta-barium-borate (BBO)\ crystals, each of $0.6$ mm depth with
perpendicular optical axis, to generate photon pairs at the wavelength of $%
800$ nm. Birefringent quartz crystals induce large shape mismatch between
horizontal and vertical polarization components of the pulses, and the input
state to the box thus has vanishing entanglement as verified by quantum
state tomography. The setup inside the box realizes an entanglement filter through a dissipative optical
channel (environment) which selects out one of the maximally entangled Bell
states as the fitted state from this environment. After the box, the
entanglement is verified through quantum state tomography, using wave
plates, single-photon detectors, and coincidence measurements.}
\end{figure}

To experimentally confirm that the input state to the optical channel has
vanishing entanglement, we perform quantum state tomography on the state $%
\rho _{in}$. For two-qubit states, the quantum state tomography is done with
$16$ independent measurements in complementary bases and the density matrix
is reconstructed using the maximum likelihood method \cite{15}. The density
matrix from the experimental measurement is shown in Fig. 2 when $%
c_{0}=c_{1}=1/\sqrt{2}$. The data is consistent with the state $\rho
_{in}=(\left\vert HV\right\rangle \left\langle HV\right\vert +\left\vert
VH\right\rangle \left\langle VH\right\vert )/2$ with small off-diagonal
matrix elements. The off-diagonal matrix elements result from the small but
non-zero overlap between the pulse shapes $f_{H}\left( t\right) $ and $%
f_{V}\left( t\right) $ for the horizontal and the vertical polarization
components. From the measured density matrix, the concurrence for the input
state is found to be $C=0.040\pm 0.024$, which indicates that the
entanglement for input state is close to zero. The error bars in $C$ and
other experimentally measured quantities account for the statistical error
associated with the photon detection under the assumption of a Poissonian
distribution for the photon counts.

\begin{figure}[tbp]
\includegraphics[width=8cm,height=8cm]{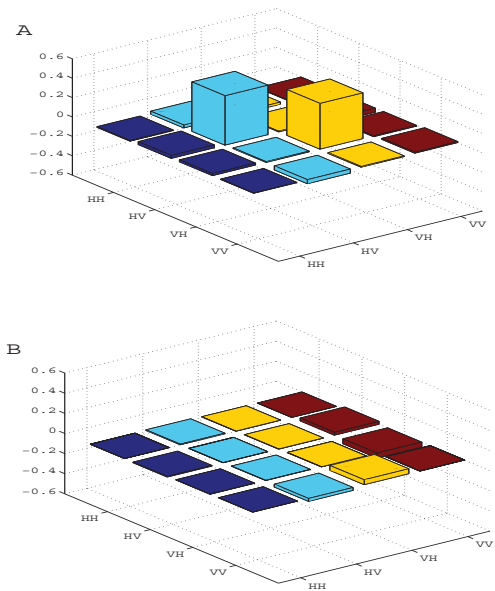}
\caption[Fig. 1 ]{Real (A) and imaginary (B) parts of the density matrix of
the input state to the environment reconstructed through the quantum state
tomography.}
\end{figure}

Now the input state goes through an entanglement filter represented by the
optical channel shown in Fig. 1 which has a tunable environmental coupling
induced by an attenuator. The attenuator, together with the Hong-Ou-Mandel
interference setup in Fig.1, introduces different decay rates for the Bell
states $\left\vert \Psi ^{+}\right\rangle $ and $\left\vert \Psi
^{-}\right\rangle $. After the first beam splitter in Fig.1, the two photons
will go along the same path (with $1/2$ probability for each path) if they
are in the state $\left\vert \Psi ^{+}\right\rangle $ and different paths if
they are in $\left\vert \Psi ^{-}\right\rangle $. An attenuator induces loss
of photons in one of the paths with the attenuation factor $e^{-\gamma }$.
So the net attenuation factors for the state $\left\vert \Psi
^{+}\right\rangle $ and $\left\vert \Psi ^{-}\right\rangle $ are given
respectively by $\left( 1+e^{-2\gamma }\right) /2$ and $e^{-\gamma }$.
Apparently, the state $\left\vert \Psi ^{+}\right\rangle $ has a smaller
attenuation in this optical channel. So, the environment favors the
entangled state $\left\vert \Psi ^{+}\right\rangle $ by killing its
competing component $\left\vert \Psi ^{-}\right\rangle $ at a faster rate.
After this optical channel, the emerging photons after the second beam
splitter is described by the effective state $\rho _{out}$ in the form of
Eq. (2) with the decay ratio $\beta =2/\left( e^{-\gamma }+e^{\gamma
}\right) $ (Note that similar to other SPDC experiments, the vacuum
components at any output port of the optical channel are dropped as they
will be erased by the single-photon detection and coincidence measurement).
The concurrence of $\rho _{out}$ is then given by $C=\tanh ^{2}(\gamma /2)$,
which approaches to the unity corresponding to a maximally entangled state
when the attenuation factor $e^{-\gamma }\ll 1$.

To verify entanglement generated by the entanglement filter through the
environmental selection, we perform quantum state tomography on the output
state $\rho _{out}$. Figure 3 shows the elements of the density matrix
reconstructed from the experimental data. The output $\rho _{out}$ is pretty
close to a pure state with the density matrix $\left\vert \Psi _{\theta
}\right\rangle \left\langle \Psi _{\theta }\right\vert $, where $\left\vert
\Psi _{\theta }\right\rangle =\left( \left\vert HV\right\rangle +e^{i\theta
}\left\vert VH\right\rangle \right) /\sqrt{2}$ with $\theta \approx 0.150\pi
$. The state deviates by a relative phase shift $\theta $ from the survivor
state $\left\vert \Psi ^{+}\right\rangle $ that one expects from the above
analysis. The reason for this deviation is that the beam splitter in our
setup has birefringence, which induces slightly different relative phases
for the horizontal and the vertical polarization components on the output
paths (see the characterization in the method section). Note that $\theta $
is just a fixed phase shift which can be easily compensated with no
influence on entanglement. From the measured density matrix $\rho _{out}$,
the entanglement fidelity for the output state, defined as the overlap of $%
\rho _{out}$\ with the closest maximally entangled state, is found to be $%
F_{e}=0.947\pm 0.019$. A bound on the entanglement fidelity with $F_{e}>1/2$
is a witness for entanglement \cite{16}. From the measured matrix elements,
the concurrence for the state $\rho _{out}$ is calculated, given by $C\left(
\rho _{out}\right) =0.902\pm 0.028$, which suggests that substantial
entanglement has been generated in our experiment through the environmental
selection.

\begin{figure}[tbp]
\includegraphics[width=8cm,height=8cm]{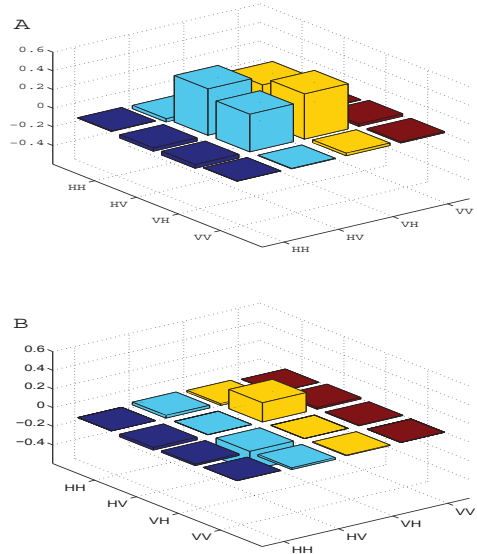}
\caption[Fig. 1 ]{Real (A) and imaginary (B) parts of the density matrix of
the output state after the environmental selection reconstructed through the
quantum state tomography..}
\end{figure}

To confirm that the environmental selection is the mechanism for
entanglement generation, we control the survival probabilities for the
states $\left\vert \Psi ^{+}\right\rangle $ and $\left\vert \Psi
^{-}\right\rangle $ in the optical channel by tuning the attenuation $\gamma
$ and measure how the entanglement varies under change of the environmental
selection strength (which corresponds to a partial entanglement filter). For
each case, the output state of the optical channel is reconstructed
experimentally through the quantum state tomography and its entanglement is
directly calculated from the measured density matrices. In Fig. 4, we show
the measured entanglement in term of concurrence for the output state under
different loss rate $\epsilon =1-e^{-\gamma }$, and the results are compared
with the theoretical prediction based on the environmental selection. With a
small loss rate $\epsilon $, which corresponds to the case of weak
environmental selection, the entanglement is tiny. As the strength of the
environmental selection increases by tuning up the loss $\epsilon $,
entanglement significantly increases and eventually approaches the unity
when $\epsilon \rightarrow 1$. The agreement between the data and the
theoretical prediction within the uncertainty imposed by the experimental
imperfection indicates that the environmental selection is the underlying
mechanism for entanglement generation.

\begin{figure}[tbp]
\includegraphics[width=8cm,height=5cm]{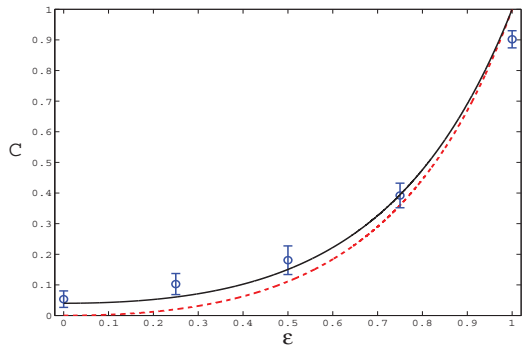}
\caption[Fig. 1 ]{The entanglement (in concurrence $C$) of the output state
as a function of the loss rate $\protect\epsilon$ induced by the attenuator.
The density matrices are reconstructed at each data point through the
quantum state tomography, and the entanglement is calculated from the
experimentally measured density matrices, with the error bar coming from the
statistical error in photon counts. The solid (dashed) lines are from
theoretical predictions based on the environmental selection mechanism. We
use the ideal input state in Eq. (1) with $|c_0|^2=1/2$ for the dashed line
(where $C=\tanh^2(\protect\gamma/2)=\protect\epsilon^2/(2-\protect\epsilon%
)^2 $) and the real input state reconstructed in Fig. 2 through quantum
state tomograph for the solid line. }
\end{figure}

The entanglement generated from the environmental selection is robust under
change of various experimental parameters or configurations, as shown in
Table 1. First, it is insensitive to the initial state. We arbitrarily
rotate the polarization of the pumping laser through a half wave plate, and
entanglement of the output state remains basically unchanged within the
experimental error bar. For instance, when the pumping pulse is set to
horizontal polarization, the input state to the optical channel becomes $%
\left\vert HV\right\rangle $ (apparently unentangled), which can be written
as a superposition of $\left\vert \Psi ^{+}\right\rangle $ and $\left\vert
\Psi ^{-}\right\rangle $ states. As the $\left\vert \Psi ^{-}\right\rangle $
component is diminished by the environmental selection, the output state
becomes a maximally entangled one. The measured concurrence in this case is $%
0.902\pm 0.020$. Second, the entanglement generation here is also
insensitive to some environmental parameters. For instance, if we insert a
quarter wave plate to change the relative phase between the two arms of the
optical channel in the box of Fig. 1, the change to the measured concurrence
shown in Table 1 is tiny and within the experimental error bar. Finally,
note that in the conventional SPDC setup, the pulse shape compensation is a
critical requirement for entanglement generation. Here, when the
entanglement is generated by the environmental selection, we do not need any
shape compensation after or before the BBO\ crystal. If we remove the
birefringent quartz crystals or add more of them in Fig. 1, the variation of
the entanglement fidelity and the concurrence remains negligible as shown in
Table 1.

\begin{tabular}{|c|c|c|}
\hline
Configurations & Entanglement fidelity & Concurrence \\ \hline
case I & $0.947 \pm 0.019$ & $0.902 \pm 0.028$ \\ \hline
case II & $0.940 \pm 0.012$ & $0.902 \pm 0.020$ \\ \hline
case III & $0.934 \pm 0.014$ & $0.897\pm 0.026$ \\ \hline
case IV & $0.952 \pm 0.013$ & $0.923 \pm 0.026$ \\ \hline
\end{tabular}%
\newline
Table 1: Entanglement fidelity and concurrence of the output state after the
environmental selection under different configurations of the input state
and environment. Case I corresponds to a mixed input state to the
environment where the density matrix is shown in Fig. 2. In case II, the
polarization of the pumping laser is in horizontal direction, corresponding
to a product input state $\left\vert HV\right\rangle $ to the environment.
In case III, we insert a quarter wave plate in one arm of the optical
channel, which induces a relative phase shift between the two polarization
components. In case IV, we remove one quartz crystal which changes the shape
mismatching for the polarization components in the input state.

In this paper, we report an experiment that demonstrates a new type of
entanglement filter based on the environmental selection, a key concept in
the recent theory of quantum Darwinism. We have experimentally confirmed
that the entanglement filter generates high-fidelity entangled states
through selection-out of the fittest component from the initially
unentangled states. The entanglement filter may prove to be a useful quantum
information device, with its ability to robustly generate and manipulate
entanglement.

\textbf{Appendix: Birefringence of the beam splitter} A balanced
non-polarizing beam splitter induces a transformation $a_{1}\rightarrow
(a_{1}+a_{2})/\sqrt{2}$, $a_{2}\rightarrow (a_{2}-a_{1})/\sqrt{2}$ to its
input-output modes for both the horizontal ($a=h$) and the vertical ($a=v$)
polarization components. However, the beam splitters used in our experiment
have a small birefringence where the transformation can be represented by $%
h_{1}\rightarrow (h_{1}+h_{2})/\sqrt{2}$ and $v_{1}\rightarrow
(v_{1}e^{i\theta _{1}}+v_{2}e^{i\theta _{2}})/\sqrt{2}$. Compared with the
conventional transformation, we should replace $\left\vert
V_{1}\right\rangle $ and $\left\vert V_{2}\right\rangle $ with $\left\vert
V_{1}\right\rangle e^{i\theta _{1}}$ and $\left\vert V_{2}\right\rangle
e^{i\theta _{2}}$, and the fittest state selected out by the environment
thus becomes $\left\vert \Psi _{\theta }\right\rangle =\left( \left\vert
HV\right\rangle +e^{i\theta }\left\vert VH\right\rangle \right) /\sqrt{2}$
with $\theta =\theta _{1}-\theta _{2}$. Using a laser beam at 800 nm
wavelength, we measure the angles $\theta _{1}$ and $\theta _{2}$ by setting
the input at the polarization state $(|H\rangle +|V\rangle )/\sqrt{2}$ and
find that $\theta _{1}=7.2^{o}$ and $\theta _{2}=-18.6^{o}$ for our beam
splitter, which gives $\theta =25.8^{o}\simeq 0.143\pi $, in good agreement
with the state in Fig. 3 reconstructed from the quantum state tomography.

\textbf{Acknowledgement} This work was supported by the National Basic
Research Program of China (973 Program) 2011CBA00300 (2011CBA00302) and the
NSFC Grant 61033001. LMD acknowledges in addition support from the IARPA
MUSIQC program, the ARO and the AFOSR MURI program.


\begin{thebibliography}{99}
\bibitem{1} W. H. Zurek, Rev. Mod. Phys. 75, 715 (2003).

\bibitem{2} W. H. Zurek, Nature Phys. 5, 181 (2009).

\bibitem{3} S. Lloyd, Nature Phys. 5, 164 (2009).

\bibitem{4} H. Ollivier, D. Poulin, W. H. Zurek, Phys. Rev. Lett. 93, 220401
(2004).

\bibitem{5} R. Blume-Kohout, W. H. Zurek, Phys. Rev. Lett. 101, 240405
(2008).

\bibitem{6} C. Monroe, Nature 416, 238 (2002).

\bibitem{6a} H. F. Hofmann, S. Takeuchi, Phys. Rev. Lett. 88, 147901 (2002).

\bibitem{6b} R. Okamoto et al., Science 323, 483 (2009).

\bibitem{7} L.-M. Duan, M. Lukin, J. I. Cirac, P. Zoller, Nature 414, 413
(2001).

\bibitem{8} D. L. Moehring et al., Nature 449, 68 (2007).

\bibitem{9} S. Diehl et al., Nature Physics 4, 878-883 (2008).

\bibitem{10} F. Verstraete, M. M. Wolf, J. I. Cirac, Nature Physics 5, 633
(2009).

\bibitem{11} H. Krauter et al., Phys. Rev. Lett. 107, 080503 (2011).

\bibitem{12} F. Bariani, Y. O. Dudin, T. A. B. Kennedy, A. Kuzmich, Phys.
Rev. Lett. 108, 030501 (2012).

\bibitem{13} W. K. Wootters, \textit{Phys. Rev. Lett.} \textbf{80}, 2245
(1998).

\bibitem{15} D. F. V. James et al., \textit{Phys. Rev. A} \textbf{64},
052312 (2001).

\bibitem{16} B. B. Blinov, D. L. Moehring, L. M. Duan, C. Monroe, Nature
428, 153 (2004).
\end{thebibliography}
\end{document}